\documentclass[12pt]{article}
\usepackage{amsmath,amssymb,amsthm,mathrsfs,url,hyperref}

\title{Some Things I Have Learned From Detlef D\"urr}
\author{
Roderich Tumulka\footnote{Fachbereich Mathematik, Eberhard-Karls-Universit\"at T\"ubingen, 
	Auf der Morgenstelle 10, 72076 T\"ubingen, Germany. 
	E-mail: roderich.tumulka@uni-tuebingen.de}
}
\date{June 27, 2023}

\newcommand{\Hilbert}{\mathscr{H}}
\newcommand{\be}{\begin{equation}}
\newcommand{\ee}{\end{equation}}
\newcommand{\scp}[2]{\langle #1|#2\rangle}

\newcommand{\PPP}{\mathbb{P}}
\newcommand{\sE}{\mathscr{E}}

\newcommand{\sZ}{\mathscr{Z}}

\begin{document}
\maketitle
\begin{abstract}
Detlef D\"urr (1951--2021) was a theoretical and mathematical physicist who worked particularly on the foundations of quantum mechanics, electromagnetism, and statistical mechanics. This piece is a rather personal look back at him and his science.
\end{abstract}

\section{Introduction}

Sadly, Detlef D\"urr passed away after short and severe illness on January 3, 2021, at the age of 69. I had the privilege to be one of his Ph.D.\ students in 1998--2001, and we continued collaborating until his death. At first, he was my supervisor, then he became my colleague and dear friend. I still find his research achievements very impressive and inspiring, and I would like to try to convey here why.

\section{An Example}

Detlef's research intertwined mathematics, physics, and philosophy. To illustrate this, I pick as an example a mathematical result Detlef published in 2004 jointly with his long-time collaborators Shelly Goldstein and Nino Zangh\`\i\ \cite{DGZ04}. I call it the main theorem about POVMs. (A POVM, or positive-operator-valued measure, is for our purposes a family of positive operators on a Hilbert space that add up to the identity operator.) Here is a (somewhat informal) statement of the theorem.

\bigskip

\noindent{\bf Theorem.} {\it Let $S$ be a quantum system with Hilbert space $\Hilbert$. For any conceivable experiment $\sE$ that can be conducted on $S$ when it has arbitrary wave function, there is a POVM $E$ on the set $\sZ$ of possible outcomes of $\sE$ acting on $\Hilbert$ such that for every $\psi\in\Hilbert$ with $\|\psi\|=1$, the probability distribution of the outcome $Z$ of $\sE$ is}
\[
\PPP_\psi(Z=z) =\scp{\psi}{E(z)|\psi}~~~~\text{for all }z\in\sZ.
\]

\bigskip

The statement is mathematics as it can be formulated rigorously and given a proof, see \cite{DGZ04} or \cite[Sec.~5.1]{Tum22}. It is physics as it concerns the physics question of which probability distributions occur in quantum experiments and how they can depend on the wave function $\psi$. It also has a flavor of philosophy (although it is an indisputable scientific fact) because of its foundational character, or even more so because it is based on an analysis of the measurement process that is usually treated as un-analyzable, but most of all because of the status it confers to the observables: In almost every textbook on quantum mechanics, observables are self-adjoint operators that enter the theory through a \emph{postulate}; but here, observables are given by POVMs, and they come in through a mathematical \emph{analysis} of the measurement process. That is a kind of radical break with the attitude prevailing in quantum mechanics: analysis instead of postulate! 

This theorem opens up a new way of thinking about quantum observables. For me, it removes the mystery about quantum observables. Indeed, in textbook quantum mechanics the observables retained a mysterious air as they were thought of as physical quantities but do not actually have values, not to speak of the fact that they do not commute. Here, observables are something different: they are mathematical objects that encode how the probability distribution of the random outcome $Z$ of an experiment depends on $\psi$. Since this dependence is quadratic, it is rather obvious that this mathematical object should be an operator $E(z)$. And operators do not commute in general. No mystery left.

The above theorem has various applications. For example, it provides a clear justification of why superselection rules hold under suitable conditions, of various versions of the no-signaling theorem, of why two probability distributions over wave functions with equal density matrices are empirically indistinguishable, and of why an experiment on one of two entangled systems has outcomes with distribution determined by the reduced density matrix.

\section{Detlef's Questions}

A key trait that makes Detlef's research findings come alive to me is that they are about unveiling how the world works: what its fundamental physical laws are, and how to explain macroscopic phenomena like randomness or the arrow of time. In other words, it impresses me that Detlef's research adds to our understanding of the world. That may almost seem impossible today. We are used to the idea that our understanding of the world is based on the works of Lavoisier and Copernicus and Einstein and other guys from previous centuries, but not on contributions from contemporary scientists. We are used to the idea that although present-day scientists do computations much more difficult than those of Lavoisier and Copernicus and Einstein, the present-day results are of lesser significance and concern details that only few specialists have ever heard of. But Detlef didn't see it that way. He saw possibilities of progress in our days on rather fundamental questions of science. 

I have learned from him that when we study scientific issues, we need to get to their bottom, that is, to reach full understanding. For example, he wanted to get to the bottom of the second law of thermodynamics (the statement that the entropy of a closed system can only increase) and understand the origin of the arrow of time. He thought it was part of a scientist's job to actually \emph{understand} her or his field, and that the quality of a researcher's results will depend on how well she or he understands which of the available theories or approaches work and why. In particular, Detlef thought it was part of his job to understand whether and how statistical mechanics explains entropy increase and the arrow of time. Since a lot of different approaches to this question have been proposed in the literature, for example some using the Gibbs entropy and some the Boltzmann entropy \cite{GLTZ20}, he felt it was important to think through which of these approaches were valid, and he inspired me to do the same. I arrived at the same conclusion as he had, not because I would repeat what he said but because he had arguments that made sense (and that was ultimately because he had honestly and seriously thought about it). The conclusion was, put very briefly, that Boltzmann's approach \cite{Bol1898}, based on typicality, provided the crucial elements. This understanding then formed the basis of Detlef's technical, mathematical work about the origin of randomness from chaos and typicality, in particular work on deriving that the velocity of a tracer particle in a classical many-body system (such as a hard sphere gas) follows approximately a Wiener process \cite{DGL81}.

\section{Quantum Mechanics}

Physics today is really in a crisis; it is a quantum crisis. In future centuries, physicists will say they cannot figure out how mainstream physicists in the 20th and 21st centuries thought quantum mechanics works because what they wrote does not actually make sense. Detlef saw this crisis clearly, and instead of shrugging his shoulders, he tried to get to the bottom of quantum mechanics. He insisted that physical theories have to make sense. I have learned from him that it is important to openly criticize orthodox views where they deserve criticism.

In fact, Detlef was frank with criticism. He criticized my work and ideas a lot. But it was the kind of criticism that belongs in a fair debate. Debates have the purpose of finding the truth, not the purpose of defending a particular person. The debates with Detlef often led to agreement, or at least partial agreement; all people involved had learned something.

Detlef had known and collaborated with Shelly since the late 1970s and with Nino since the mid  1980s. The shared goal to get to the bottom of quantum mechanics led to their intensified joint collaboration on that subject in the late 1980s and to a series of
D\"urr-Goldstein-Zangh\`\i\ (DGZ) papers since about 1990, most of them collected later in Ref.\ \cite{DGZ13}. They jointly arrived at the conclusion that, for quantum mechanics, Bohm's approach \cite{Bohm52a,DT09} provided the crucial elements. They clarified Bohm's theory, defined it in more coherent way than Bohm himself, and derived and motivated it in a more direct way; one could say they put it upside up. Detlef coined the name ``Bohmian mechanics'' in analogy to Newtonian mechanics. By combining the approaches of Bohm and Boltzmann, DGZ clarified the status of the Born rule in Bohmian mechanics as based on typicality \cite{DGZ92}, thereby removing the need for an approach to equilibrium through mixing. 

Bohmian mechanics is not just a counter-example to some orthodox claims, it is the most serious theory of quantum mechanics that we have. 
By using a clear theory like this, we can clear up many confusing issues such as contextuality or the status of observables; we can see more clearly which general statements or proofs are relevant (such as the main theorem about POVMs); we can see more clearly what the problems are with various fields of quantum physics (such as quantum field theory) and take concrete steps to make progress on these problems. For example, quantum field theory involves the creation and annihilation of particles, which leads to the question of how to incorporate particle creation in Bohmian mechanics. I had the privilege to work with DGZ on the development of such extensions.

I also think it was important that DGZ took seriously the possibility that our relativistic space-time might have a preferred foliation (slicing) into spacelike hypersurfaces, even though this might at first seem against the spirit of relativity. Equations for an adaptation of Bohmian mechanics using a preferred foliation were worked out in 1999 \cite{HBD} (actually, in a general-relativistic setting, although the published paper limits its discussion to the special-relativistic case). Since Bell's theorem \cite{Bell87} shows that nonlocality is inevitable, the nonlocality of Bohmian mechanics is good, not bad.

At the same time, Detlef was not at all dogmatic about Bohmian mechanics. For example, he supported (and contributed himself \cite{DHK11,BDH13} to) research about a competitor to Bohmian mechanics, the Ghirardi-Rimini-Weber theory of wave function collapse. 
It wasn't that he was dissatisfied with Bohmian mechanics and searched for alternatives. Rather, he felt that theories of quantum mechanics that actually make sense, that describe a coherent picture of reality, should be explored.

\section{Mathematics}

Some mathematicians tend to exaggerate the importance of proofs; they have a low opinion of the person who came up with a conjecture and a high opinion of the person who proved it, even if the proof did not introduce any particularly new ideas but arose mainly from persistence and diligence. Detlef was not like that. Although he was a mathematician and belonged to a math department, he had a deep appreciation for non-rigorous considerations. But that doesn't mean that he was imprecise or didn't do proofs. On the contrary, he encouraged me and other students to provide rigorous studies of the physical concepts we considered, in part because that forces us to be careful and precise. 

Correspondingly, he himself collaborated in a large number of such studies: for example, apart from the theorem mentioned initially \cite{DGZ04}, he contributed to mathematical proofs of global existence of Bohmian trajectories \cite{bmex}, that Bohmian trajectories become straight lines in the scattering limit \cite{RDM}, that the probability flux (and thus the Bohmian arrival places) agree with the scattering matrix \cite{DGMZ}, 
and to constructing the unitary time evolution of the quantized Dirac field in an external electromagnetic field \cite{DDMS}. He taught through his example how careful mathematical study should complement physical theory, and conversely how physics and other real-world applications give us orientation in a sea of mathematical possibilities. I believe that if you feed students with unmotivated mathematics, then it reduces their ability to think about mathematics.

Here is another thing that, although it is kind of obvious, I had never realized until Detlef pointed it out to me: classical electrodynamics is inconsistent. After all, the Lorentz force law requires that we evaluate the electromagnetic field $F_{\mu\nu}$ at the location of the charged particle, but exactly there it diverges as a consequence of the Maxwell equation. Textbooks on classical electrodynamics had never mentioned this basic fact.
Going further, for solving the ultraviolet divergence problem of quantum electrodynamics, it might be a good start to try to solve it for classical electrodynamics. This has inspired Detlef's interest in Wheeler-Feynman electrodynamics \cite{BDD13} and in shape dynamics \cite{DGZ20}, two approaches that might offer ways out and that he published mathematical studies of.

\section{Directions}

Let me come back to Detlef's choice of research problems. It was motivated by the goal of understanding nature. It was not motivated by seeking applause, or trying to win (explicit or implicit) competitions against other scientists. People who see science as a competition between the smartest minds and wish to score highly have a tendency to choose problems that are popular (that many others have already worked on) and whose answer will be uncontroversial; they invest their time and energy in improving known results (providing, say, more accurate approximations and tighter bounds) and hope for recognition. Competition tends to remove the sense of intrinsic value: people who believe that the main reason for practicing the violin, or math, or swimming is to be better than others will also believe that nobody would play the violin, or think about math, or swim just for the joy of it. Competitions are usually competitions in doing something useless. Detlef's style was different. It was not that he didn't enjoy recognition, or that he didn't care what other people said. But on matters on which he could judge for himself, he did judge for himself, and didn't follow widespread opinion when he knew it was flawed. I think that research problems that will win competition and applause today but have little intrinsic value will be of little interest to future generations of scientists. As I am writing this, I have to think of a quotation of Albert Einstein \cite{Ein55}:
\begin{quotation}
``Try not to become a man of success, but rather try to become a man of value.''
\end{quotation}
This fits with what I described if the ``man of success'' is the scientist who primarily wants to gain applause for his work regardless of its actual value while the ``man of value'' primarily wants to do work of lasting scientific value regardless of how much applause it may win or not. I think that Detlef was a man of value.

\end{document}